\documentclass[usenatbib]{mnras}
\usepackage[T1]{fontenc}

\DeclareRobustCommand{\VAN}[3]{#2}
\let\VANthebibliography\thebibliography
\def\thebibliography{\DeclareRobustCommand{\VAN}[3]{##3}\VANthebibliography}

\usepackage{graphicx}	% Including figure files
\usepackage{amsmath}	% Advanced maths commands
\usepackage{amssymb}	% Extra maths symbols
\usepackage{ulem}

\newcommand{\accunit}{\,{\rm km^2\,s^{-2}\,kpc^{-1}}}	
\newcommand{\accunitperkpc}{\,{\rm km^2\,s^{-2}\,kpc^{-2}}}	
\newcommand{\opunit}{\,{\rm km\,s^{-1}\,kpc^{-1}}}	
\newcommand{\potunit}{\,{\rm km^2\,s^{-2}}}

\title[Non-equilibrium dynamics]{Non-equilibrium in the Solar Neighbourhood using dynamical modelling with Gaia DR2 }

\author[Kipper et al]{
Rain Kipper,$^{1}$\thanks{E-mail: rain.kipper@ut.ee}
Peeter Tenjes,$^{1}$
Elmo Tempel,$^{1}$
Roberto de Propris$^{2}$
\\
$^{1}$Tartu Observatory, University of Tartu, Observatooriumi 1, 61602 T\~oravere, Estonia\\    
$^{2}$Finnish Centre for Astronomy with ESO (FINCA), University of Turku, Vesilinnantie 5, 20014, Turku, Finland\\
}

\date{Accepted XXX. Received YYY; in original form ZZZ}

\pubyear{2015}

\begin{document}
\label{firstpage}
\pagerange{\pageref{firstpage}--\pageref{lastpage}}
\maketitle

\begin{abstract}
Matter distribution models of the Milky Way galaxy are usually stationary, although there are known to be wave-like perturbations in the disc at $\sim10\%$ level of the total density. Modelling of the overall acceleration field by allowing non-equilibrium is a complicated task. We must learn to distinguish whether density enhancements are persistent or not by their nature. In the present paper, we elaborate our orbital arc method to include the effects of massless perturbations and non-stationarities in the modelling. The method is tested by modelling of simulation data and shown to be valid. We apply the method to the Gaia DR~2 data within a region of $\sim 0.5$~kpc from the Sun and confirm that acceleration field in the Solar Neighbourhood has a perturbed nature -- the phase space density along the orbits of stars grow in the order of $h\lesssim 5\%$ per Myr due to non-stationarity. This result is a temporally local value and can be used only within the timeframe of a few Myrs.  An attempt to pinpoint the origin of the perturbation shows that the stars having larger absolute angular momentum  are the main carriers of the local perturbation. As they are faster than the average thin disc star, they are either originating further away and are close in their pericentre or they are perturbed locally by a fast co-moving perturber, such as gas disc inhomogenities. 
\end{abstract}


%%%%%%%%%%%%%%%%%%%%%%%%%%%%%%%%%%%%%%%%%%%%%%%%%%

%%%%%%%%%%%%%%%%% BODY OF PAPER %%%%%%%%%%%%%%%%%%

\section{Introduction}
Although in general galaxies are not isolated in space and are thus, dynamically interrelated, their mass distribution models are usually stationary (see, e.g. a review by \citealt{Rix:2013}) or related to some specific aspect of galaxy, e.g. phase-space spirals in \citet{widmark2021weighing}. The stationarity assumption is a widely used approximation as the periods of stellar oscillations in a typical galaxy are less than a Gyr. Stationary models often allow us to  describe the internal velocity distribution quite satisfactorily \citep{Cappellari:2013, Leung:2018, Zhu:2018}.

Although stationary modelling can provide satisfactory results in modelling, there are slow changes in discs shaped by minor mergers \citep{Abadi:2003, Villalobos:2008, Moster:2010, Bird:2012}. Stellar halos may be formed as due to disruption of several tens of dwarf galaxies \citep{Johnston:2008}, assembly of the dark halo from subhalos created fluctuating overall potential influencing visible components of galaxies \citep{Carlberg:2019, Vasiliev:2020}. In addition to these minor events in case of the Milky Way (MW) galaxy there have been in the past by at least one significant merger \citep{Helmi:2018}. 
Although most of these ancient marks of the mergers are smoothed out by now, the traces remain still visible in action space. 

Recent observations give a strong hint that there are time-dependent perturbations of the density in our Galaxy and conventional dynamics modelling need improving \citep{Haines:2019, Salomon:2020}. Systematic vertical motions in the stellar disc and north-south asymmetry were found by \citet{Widrow:2012, Williams:2013, Carlin:2013, Yanny:2013}, Gaia satellite has also revealed a rather complex structure of the Milky Way galaxy in 6D phase space. For example, \citet{Ramos:2018} found from Gaia DR2 data that the phases of stars are not well mixed for all orbits. \citet{Antoja:2018, Kawata:2018, Trick:2019} also found similar arcs and ridges. 

Observations of systematic vertical motions imply that MW disc is undergoing compression and expansion perpendicular to the plane \citep{Banik:2017}. These perturbations can be caused by the passages of globular clusters, dwarf galaxies, dark matter subhalos, passing spiral arms etc.

The main obstacle regarding the modelling of perturbed systems arise from collisionless Boltzmann equation 
\begin{equation}
    \frac{\partial f}{\partial t} + \sum_{i=1}^3
    \frac{\partial f}{\partial v_i}\frac{ {\rm d}v_i}{ {\rm d}t} +
    \frac{\partial f}{\partial x_i}\frac{ {\rm d}x_i}{ {\rm d}t} = 0.
    \label{eq:Boltzmann_classical}
\end{equation}
Here $f$ is the phase space density and $i$ indexes components of velocity and position vectors $\mathbf{v} $ and $\mathbf{x}$.
It contains one observable variable (phase space density) and its spatial derivatives. But it also contains a time derivative, which is not directly observable, and accelerations (gravitational potential gradient components) that are also not observable. Hence there is only one equation and two unknowns, making the equation unsolvable. The presence of dark matter and too dim objects in mass distribution do not also allow the Poisson equation to constrain the acceleration field.

Time derivatives indicate that it would be correct to handle the potential as a function of time $\Phi (\mathbf{x}, t)$ and it may change noticeably model results derived within a stationary model. It is also necessary to estimate how reasonable the assumption of a stationary potential is or within what timescale one can handle integrals of motions to be constant for a particular stellar orbit. For example, by modelling the Milky Way bar with Gaia DR2 data \citet{Kipper:2020} estimated that due to torque from the central bar the angular momentum of the Sun $L_z$ might change within one orbital period about 30 per cent. It indicates that in barred galaxies $L_z$ is not as good integral of motion as one would like it to be. Conventional models should be handled as an approximation only. Slow changes of the angular momentum $L_z$ and radial action $J_R$ were derived from N body simulations by \citet{Wozniak:2020}, although they derived slower diffusion of these values.

Modelling of individual galaxies via N-body simulations (e.g. made to measure) is in some sense free of a stationarity assumption, but these models are rather time-consuming, and their spatial and mass resolution is yet moderate \citep{Syer:1996, Delorenzi:2007, Long:2010, Zhu:2014}.

In our earlier paper \citep{Kipper:2019} we presented a method to calculate gravitational potential derivatives (accelerations) in sufficiently small regions of a galaxy if one knows phase-space coordinates of a large number of stars in these regions.  In a later paper \citep{Kipper:2020} the method was applied for the Milky Way (MW) galaxy in the Solar neighbourhood (SN) using Gaia DR2 data. The model included a non-axisymmetric central bar but assumed stationarity. In the present paper, we improve the method by allowing non-stationarity, i.e. that the phase density is a function of time. As the observational data we use Gaia DR2 phase space coordinates.

It is not our aim here to describe the secular evolution of galaxies, although, in principle, it is not excluded. Although a perturbation can last for an extended period, we model only  a short time interval. This interval is determined both by the size of the studied region and the average speed of stars in that region. Extrapolation of the derived results for longer timescales than few Myrs is not justified. To emphasize this short-timed characterization, we  mainly use the word perturbation to describe the system's time evolution / non-stationarity. The selected small region of the MW is in the SN and used stars have all six phase-space coordinates from the Gaia DR2.

In Section~\ref{sec:model_update} we characterise the advancements we did in modelling compared to \citet{Kipper:2020} in more detailed, Section~\ref{sec:mock} checks the validity of the improvements with mock data. Thereafter, in Section~\ref{sec:gaia_fit} we present what we learned about the SN using this approach. The paper ends with Discussion (Section~\ref{sec:discussion}) and Conclusions (Section~\ref{sec:summary}).

%----------------------------_
\section{Advancements of the method}\label{sec:model_update}
The improvements of the method presented by \citet{Kipper:2020} are related to both dynamics and statistics. We describe them in subsequent sections. However, we provide a concise recap of the method we build our improvements. 

The orbital arc method developed in \citet{Kipper:2020} relies on seeking of consistency between assumptions and data. The data in our case is the distribution of stars in a specific region. First, we start selecting observed positions and velocities of a localised sample of stars and consider them as initial conditions for orbit calculations. We take the initial conditions, assume an acceleration with some free parameters and calculate small arcs of orbits for all the stars within the selected region. Coefficients in the acceleration field are found by demanding that the acceleration field in orbit calculations would distort the phase space distribution of stars in the least way compared to the phase space distribution constructed from initial conditions (observed values). By distorting we mean a changing the phase space by re-positioning a star infinitely many times along its orbital arc, or more precisely, we smooth the star over its short piece of orbit \textit{uniformly in time}. This procedure assumes stationarity. If the system is stationary, then the smoothed distribution and the data distribution would match, i.e. the likelihood for the statistical inference found by evaluating initial conditions of the same stars with the distorted phase-space distribution would be maximal. The inference gives us the acceleration field that best matches the stationarity of the data distribution. 

\subsection{Improved description of kinematics}\label{sec:meth_general_impr}
Modeling the distribution and kinematics of stars in Solar Neighbourhood's (SN) \citet{Bennett_2018} found that on top of the smooth distribution there are also wavy motions with the amplitude of about $10\%$. Hence, about $10\%$ of the stellar density in the SN has perturbed nature.
If we intend to model the MW with an accuracy greater than this, we must find a way to cope with these perturbations. At first sight, the easiest way is to model the MW by using data that are distributed over a much larger area than the SN and average out these perturbations. The smoothed distribution can be modelled using tools designed to model stable and stationary systems. This would provide a global MW model, and on top of this model, a perturbation model can be constructed. A drawback of it is the necessity to have a large amount of data over a large region, and the unknown perturbations might be too large-scale to average out (e.g. the bar). Also, it hinders to find the details of the calculated acceleration field: recovered details cannot be finer than the scale that the data were averaged out. In case of some form of perturbations, such as bar, the system can be modelled in a rotating frame to avoid the perturbations and time dependence, but it introduces an additional free parameter. 

This section will augment the orbital arc method \citep{Kipper:2019, Kipper:2020} to include the modelling of a region in the Galaxy being in a perturbed state. In this paper, we only aim to model perturbations that do not contain a significant amount of mass and are dominantly density perturbations leaving potential and acceleration field untouched. This approach is justified as the potential is essentially determined by a large amount of matter distributed over all galaxy, not locally\footnote{As $\nabla^2\Phi=4\pi G\rho$ density fluctuations determine only the second derivative of the potential fluctuations.}.  

The oPDF formalism from \citet{Han:2016} demonstrated that in a stationary galaxy the relation between a time interval (${\rm d}t$) and probability of finding a star in corresponding orbital segment (${\rm d}P$) is 
\begin{equation}
    {\rm d}P\propto{\rm d}t. \label{eq:opdf_stat}
\end{equation} 
If we intend to model a galaxy in a perturbed state, this simple proportionality does not hold anymore and if we want to use this kind of relation, we should add a perturbation term in order to cope with it:
\begin{equation}
    {\rm d}P \propto \,[1 + h'(t)]{\rm d}t\,\label{eq:dt_modification}. 
\end{equation}
Here, the $h'(t)$ term describes the perturbation of the dynamical state of a galaxy. Yet, adding a random term to Eq.~\eqref{eq:dt_modification} does not make it more correct or more apt to describe the underlying structure of a galaxy; so why would it work?

Let us imagine a galaxy consisting of only one densely populated orbit (i.e. a stellar stream like entity). We assume the stars are positioned along the orbit according to Eq.~\eqref{eq:opdf_stat}. Initial conditions and acceleration field fully determine orbit's shape. In case of orbital arc method, we may take any star from the orbit and infer the underlying acceleration field by matching proposed orbits with the rest of the stars in our hypothetical galaxy. As the orbit was populated with the assumption of stationarity, the matching required by the orbital arc method is expected to be perfect. 

\begin{figure}
    \centering
    \includegraphics{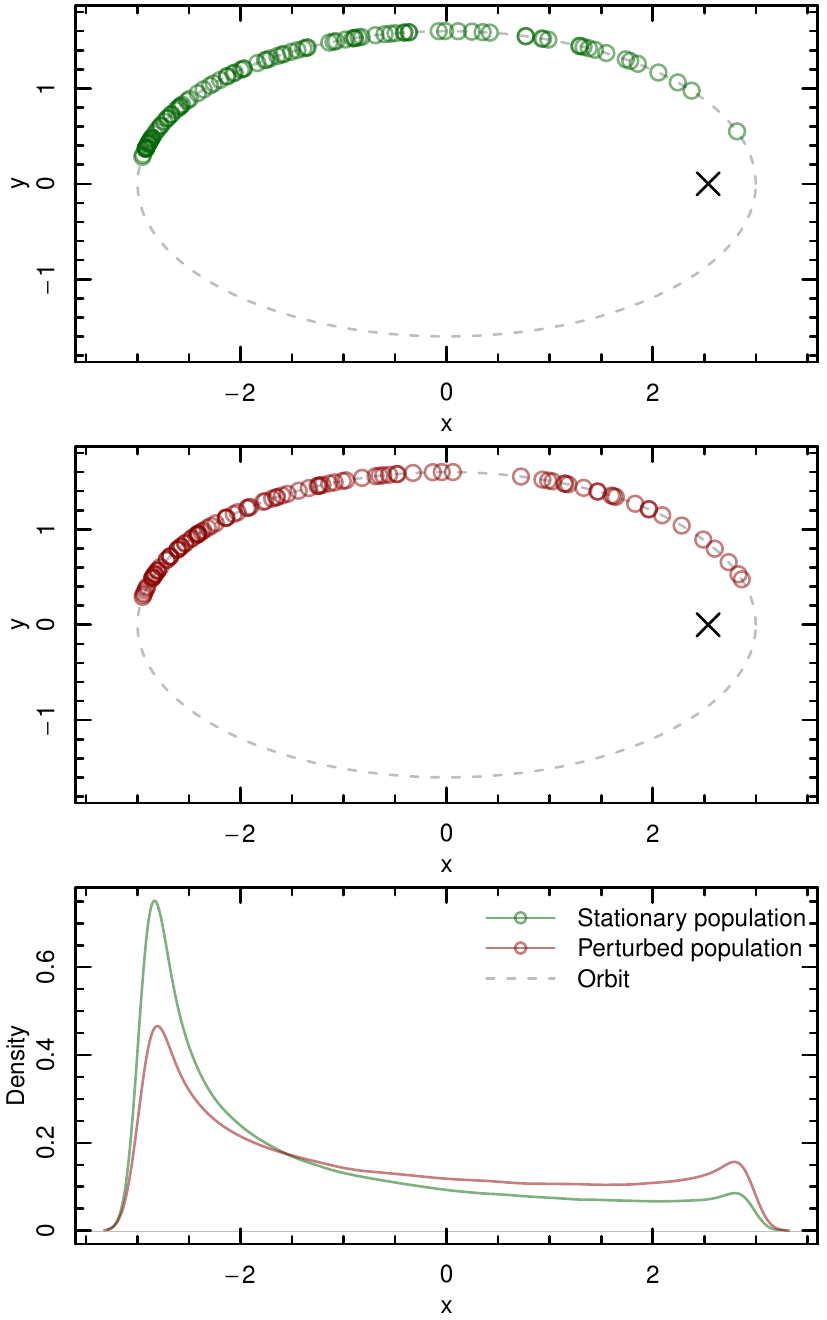}
    \caption{An illustration of the modification to the orbital arc method to model non-stationarities. The top panel shows the distribution of Keplerian orbit when the stars are populated in it according to oPDF in Eq.~\eqref{eq:opdf_stat}. The black cross shows the focal point of the Keplerian orbit. The middle panel shows the distribution of stars positioned non-stationarily. The bottom panel shows the distributions of their x-coordinates, which do not match. The update to the method has the basis that the stellar initial conditions and acceleration provide identical orbit shape. However non-stationarity affects phase space/population along the orbit without affecting velocity gradients. }
    \label{fig:method_illustration}
\end{figure}

Let us keep this hypothetical galaxy, but let us populate the orbit according to a perturbed/non-stationary state: some parts of the orbit have an elevated number density of stars. Again, orbital arc calculations can reproduce the shape of the orbit similarly well as the neighbouring points have the same kinematics as the previous case. The only differences are different number densities along the orbit. Thus, the number density of stars along an orbit depends both on the unperturbed velocities along the orbit and on the perturbations. An illustration in Fig.~\ref{fig:method_illustration} shows this concept visually. 

The stellar part of a galaxy contains a multitude of orbits overlapped with each other. Even in case of summing over all the orbits in our selected region, the shapes of orbits give us the velocity gradients in the region -- the orbit shapes determine ${\rm d}{\bf v}/{\rm d}{\bf x}$ and due to fixed initial conditions ${\rm d}f/{\rm d}{\bf v}$ is fully determined by acceleration field too. The non-stationarity does not influence the shapes of the orbits, but their population density or ${\rm d}f/{\rm d}{\bf x}$ directly while not affecting velocity gradients. The orbital arc is built to minimise changes to phase space density, hence also to keep their gradients unchanged. We conclude that the reconstruction of phase space density has different basis when we consider the sources of non-stationarity (influences only ${\rm d}f/{\rm d}{\bf x}$) and gravitational potential (influences ${\rm d}f/{\rm d}{\bf v}$ directly and ${\rm d}f/{\rm d}{\bf x}$ indirectly). A test with a simulation shows that it is a viable approach.

% ----

A more formal way to introduce this extra term to the oPDF in Eq.~\eqref{eq:dt_modification} comes from the collisionless Boltzmann equation. We start from and follow the work of \citet{Han:2016}. The classical Boltzmann equation \eqref{eq:Boltzmann_classical} written using the integrals of motion is
\begin{equation}
    \frac{\partial f}{\partial t} + \sum_{k=1}^5,
    \frac{\partial (f\dot Q_k)}{\partial Q_k}+
    \frac{\partial (f\dot\lambda)}{\partial \lambda} = 0.
    \label{eq:Boltzmann_iom}
\end{equation}
Here, the $Q_k$ characterises integrals or constants of motion and $\lambda$ is the affine parameter characterising position on the orbit. The parameter $\lambda$ need not to have a specific interpretation -- it is neccessary only that in order to determine uniquely position on the orbit, $\lambda$ must be growing monotonically along the orbit. 

The definition of integral of motion means $\dot Q_k=0$ and thus equation \eqref{eq:Boltzmann_iom} is simply
\begin{equation}
    \frac{\partial f}{\partial t} + \frac{\partial (f\dot\lambda)}{\partial \lambda} = 0. \label{eq:der_assumption}
\end{equation}
The assumption ${\partial f}/{\partial t} = 0$ leads to stationarity, which is not our aim at present. The simplest assumption not leading to stationarity is that ${\partial f}/{\partial t} \equiv {\rm const} = A$ (assumption of a constant is only our first attempt here). For example, this form can describe an approaching overdensity toward a studied region in rotation dominated disc.  
Let us make this substitution and integrate Eq.~\eqref{eq:der_assumption} with respect to $\lambda$:
\begin{equation}
    A\lambda + f\dot\lambda = B, 
\end{equation}
where $B$ is the integration constant.  
Solving the latter for $f$ gives
\begin{equation}
    f = (B - A\lambda) \frac{{\rm d}t}{{\rm d}\lambda}.\label{eq:der_f}
\end{equation}

The projection of phase space density while fixing the integral of motions provides the probability density of finding a star along the orbit. Following the approach of \citet{Han:2016} again and use \eqref{eq:der_f}, we find 
\begin{eqnarray}
    \frac{{\rm d}P}{{\rm d}\lambda}|Q \propto f(Q,\lambda) = (B - A\lambda) \frac{{\rm d}t}{{\rm d}\lambda}\label{eq:h_long_eq}.
\end{eqnarray}
In the left side $|Q$ means for a given $Q$ value. Multiplying by ${\rm d}\lambda$ gives time span star spends in a segment of an orbit. When characterizing the position along the orbit with time\footnote{Other types of characterizations are possible, but in order to infer time evolution we selected time as the affine parameter.}
we may substitute $\lambda \rightarrow t$. Writing instead of $(-A/B)\rightarrow h$ we have a similar expression to \eqref{eq:dt_modification} with $h'(t) = ht$ 
\begin{equation}
    {\rm d}P \propto (1 + ht) {{\rm d}t} \label{eq:dP_const_used}. 
\end{equation}
The practical application requires normalisation of probability, hence we substituted the equality to proportionality. The Eq.~\eqref{eq:dP_const_used} is derived for an orbit i.e. $h = h|Q$. \sout{In most cases, we make an assumption that the same level of non-stationarity applies to all orbits and neglect this dependence. We use ${\rm d}P$ in this form to model simulation and Gaia data.} For applications we \sout{also try} use an approach where the $h$ depends only on one integral of motion: $L_z$. We have implicitly assumed that $t=0$ is the time of observations.

To give an interpretation of $h$ and the perturbedness, let us look at Eq.~\eqref{eq:der_f}. Averaging Eq.~\eqref{eq:der_f} over the studied region, we have $\langle f \rangle = \langle B \rangle - \langle At \rangle = B $ as the time-dependence averages out. The time-depencence averages out since $t=0$ is defined as the time of Gaia observations and deviations from it are symmetrical (on average star can move forward and backward the same amount before exiting the studied region) and cancel out. The term $A$ was defined $A \equiv \dot f$, or $\langle A \rangle = \langle \dot f \rangle$. By normalising the intercept term in Eq.\eqref{eq:h_long_eq} to one (required by normalisation of probability) gives the interpretation of $h$ to be $h \approx \langle \dot f \rangle / \langle f \rangle$ or the averaged relative perturbedness.

\subsection{Improving statistical description}

The \textit{orbital arc} method is based on the assumption that the overall probability to find an observed point at position $p({\bf q}){\rm d}{\bf q}$ comes from the sum of $i$ orbital densities; each described by probability to find a star in the orbital point $p_i[{\bf q}(\lambda) | \vartheta]$. We denote ${\bf q}$ as an element of phase space density,  $\lambda$ characterises position along the orbit and $\vartheta$ the acceleration field parameters. The underlying acceleration parameters are determined by maximising the likelihood function with respect to the $\vartheta$
\begin{equation}
    \mathcal{L'}_g = \prod\limits_{j \in S}p({\bf q_j}) = \prod\limits_{j \in S} Z^{-1}\sum\limits_{i} p_i[{\bf q}|\vartheta]K_g({\bf q_j,q}).\label{eq:p_orb_sum}
\end{equation}
Here the $S$ is a set of stars in the region indexed by $j$. $Z$ is the normalising constant, $K_g$ is the kernel converting orbital arcs into a smooth probability distribution. More details are elaborated in \citet{Kipper:2019}. 

The paper \citet{Kipper:2020} had a drawback in statistical modelling that  having a grid-based approximation of probability produced slightly distorted $p$ values. In order to estimate and suppress it, we made multiple modellings and averaged the posterior distributions. We improve upon this by making multiple grids and $p$ estimations, and each likelihood evaluation we average the likelihoods i.e. 
\begin{equation}
    \mathcal{L} = \sum\limits_{g=1}^G\mathcal{L'}_g. \label{eq:G_variable}
\end{equation}
Here $\mathcal{L'}_g$ is a likelihood function defined by Eq.~\eqref{eq:p_orb_sum} but implemented with different grids in kernel $K$. The $g$ indexes the likelihoods, and $G$ is the number of grids averaged per likelihood evaluation. A more thorough description is provided in \citet{Kipper_in_prep} in the context of modelling galaxy evolution. 

The second improvement comes from the necessity of independence that the likelihood evaluation point cannot be included in the calculation of $p$. This is reached by requiring 
\begin{eqnarray}
    i\ne j
\end{eqnarray}
in the summing over $i$ in Eq.~\eqref{eq:p_orb_sum}. In case the number of stars/point approach infinity, this does not produce any practical bias, but in the case where the number of stars is lower, it might. A test on gaia data showed that this effect is detectable and significant when the modelling is done with a fine grid. 

In the next section, we make a set of tests on a simulation. These tests showed that these modifications produced unbiased estimator for the acceleration field parameters. 

%----------------------------_

\section{Validating the improved method on a simulation}\label{sec:mock}

Applying the method on a mock or a simulation gives us a possibility to test that the underlying method and its implementation are robust and correct. For the testing, we chose the simulation of an MW like galaxy from \citet{Garbari_2011}. The aim is to test the assumption of non-stationarity and modifications made to improve its statistical implementation. \sout{The latter can be tested easily, but for testing non-stationarity, there is a slight inconsistency. We assumed that the perturbations were nearly massless as they do not affect potential, but some of the testing regions do not fulfil it in this simulation. Bar regions have perturbations with significant mass. Even though  perturbations have significant mass in some regions, we show that the recovery of accelerations still improves when accounting non-stationarity of density. }

\subsection{The mock data}\label{sec:mock_data}

\begin{figure}
    \centering
    \includegraphics{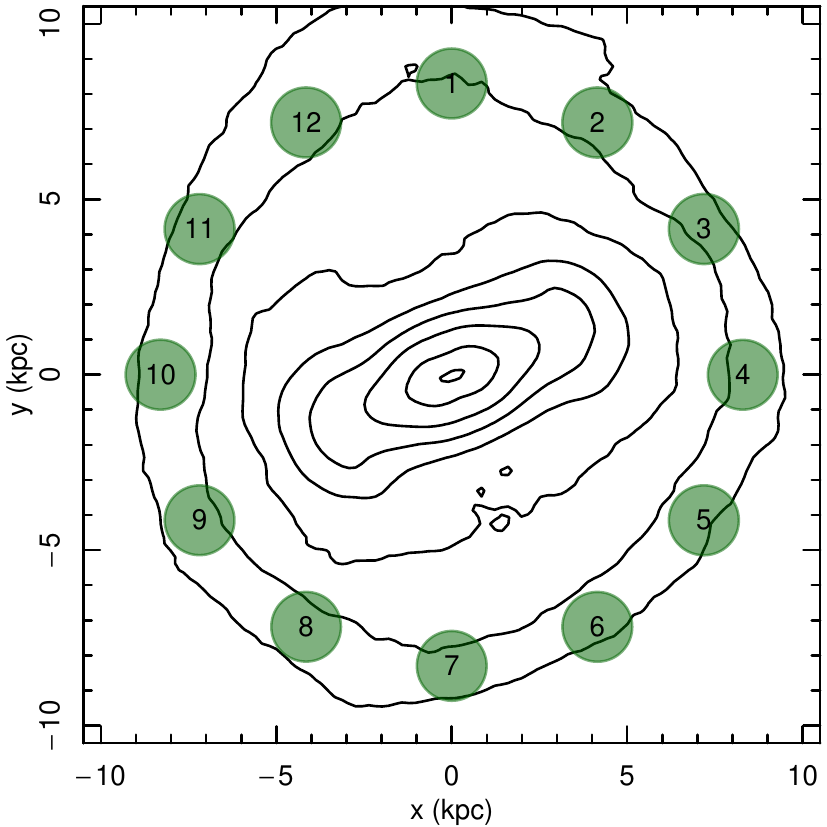}
    \caption{Sizes and positions of selected regions (green circles) for a simulated galaxy from \citep{Garbari_2011} together with isodensity contours. }
    \label{fig:mock_illustration}
\end{figure}

We selected 12 small spheroidal regions to model: all of them at the distance of 8 kpc from the galactic centre. Here and onward we denote the regions as Solar Region (SR). This designation is usually used with the corresponding number of a region. In Fig.~\ref{fig:mock_illustration} positions of these regions together with isodensity contours of the simulated galaxy are shown. 
SR regions have half-height of $0.35$ and radius of $1.0$~kpc. SRs are larger to compensate lower number density of stars in these regions, which otherwise would reduce the precision of modelling. 

We calculated the pattern speeds for $R = 1.5 - 4.5$~kpc of the simulated galaxy by solving one of the intermediate results in Tremaine-Weinberg method derivation \citep{TW:1984}
\begin{equation}
    \frac{\partial}{\partial x}
    (\Sigma  v_x) +
    \frac{\partial}{\partial y}
    (\Sigma  v_y) +
    \Omega_{\rm p}y \frac{\partial\Sigma}{\partial x} -
    \Omega_{\rm p}x \frac{\partial\Sigma}{\partial y} = 0.
\end{equation}
Here, $\Sigma$ is the surface density, $x,y$ are Cartesian coordinates, and $\Omega_{\rm p}$ is the pattern speed. 
Pattern speed in this bar region is rather constant at the value of $\Omega_{\rm p} = 37.0^{+6.0}_{-5.4}$~km\,s$^{-1}\,$kpc$^{-1}$. The uncertainty of $20\%$ is expected for the Tremaine-Weinberg method due to fluctuating nature \citep{hilmi2020fluctuations}.

\subsection{Data sampling and systematic uncertainties}
\subsubsection{Sampling noise}
For the method testing, the mock data should mimic the subsequent application as adequately as possible. The main differences with observations are that mock data are not limited by object brightnesses, the data have no uncertainties, they have perfect selection function and significantly smaller number density of stars. The brightness limit aspect was tested in \citet{Kipper:2020} and is not covered here. The observational uncertainties are minimal as we constructed the observational dataset using a very close set of stars (maximum distance is about half of a kiloparsec). 

\begin{figure}
    \centering
    \includegraphics{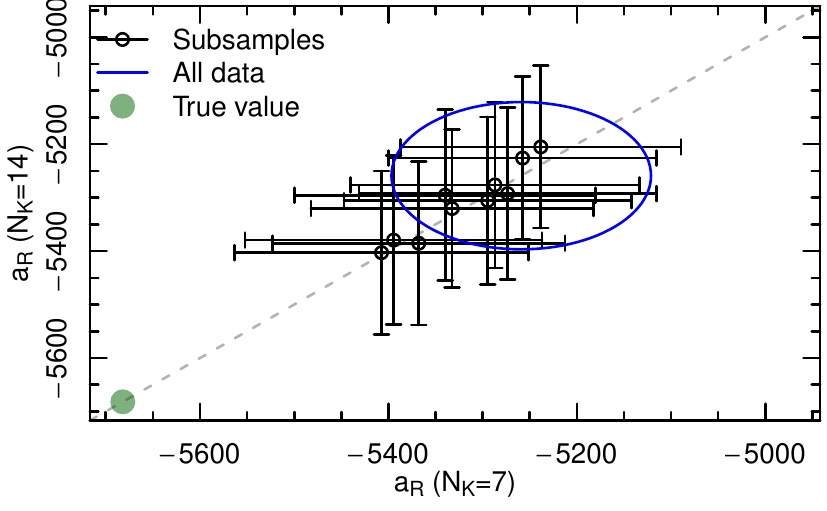}
    \caption{The sensitivity of the recovered acceleration to the sampling. Ten subsamples of SR1 each containing $75\%$ of the total sample are modelled and calculated acceleration values are presented with black colour. The blue circle shows the modelling results of the total sample and the green point the actual values from the gravitational potential. The x-axis is the recovered acceleration with coarser kernel function $K_g$ having $N_{\rm k}^2$ histogram bins. In the y-axis is the finer grid results. As neither the uncertainty nor error do not depend on the gridding, we conclude that the grid $K$ details do not influence modelled results.   }
    \label{fig:subsampling_effect}
\end{figure}

Thus, we test here only the sensitivity of the modeling due to the smaller number of stars compared to the observational application sample. This is essential as the model's accuracy is limited with the number of stars to construct the PDF in Eq.~\eqref{eq:p_orb_sum}. 
An average SR contain about $60\,000$ stars. This is significantly less than the observational sample of $400\,000$ stars. We point out that we chose the larger region in simulation than in observational application to elevate the number of stars to be more comparable. To test how sensitive the method is to reducing the size of the sample, we modelled $10$ subsamples of the SR1, each having $25\%$ less stars than the parent sample. The test results are in Fig.~\ref{fig:subsampling_effect}. It is seen a significant change of calculated $a_R$ values for different subsamples. One notice also that all of them differ within uncertainties from the actual value. As all other possible causes have been checked and have smaller influence we conclude that the reason is sampling. This is confirmed also by looking at the top panel of Fig.~\ref{fig:sim_pot_timedep} where we can see significant high-frequency noise in acceleration, being also a side-effect of sampling. In real observations the number density of stars is about an order of magnitude larger and one may  expect the sampling effects to be significantly smaller. In any case, we cannot leave this discrepancy unattended and consider these differences as the cause of systematic uncertainty described in the next section. 

\subsubsection{Systematic uncertainties}\label{sec:mock_sys_uncert}
In the previous section, we concluded that sampling could cause an offset between actual values and the modelled ones. To cover the possible offset, we describe it as systematic uncertainties (see Fig.~\ref{fig:subsampling_effect}). 

Let us consider an abstract one-dimensional statistical inference problem modelled with the Markov chain Monte Carlo (MCMC) approach \citep{MN1, MN2, MN3} similar to the \textit{Multinest} we use in present paper. Let us have a point $x$ and its proposed change to $x+\Delta x$. The change is accepted in Monte Carlo based on the log likelihood differences of these points
\begin{equation}
    \Delta \log \mathcal{L} = -\frac12\frac{x^2}{\sigma_{\rm stat}^2} + \frac12\frac{(x+\Delta x)^2}{\sigma_{\rm stat}^2} = \frac12\frac{x\Delta x + \Delta x^2}{\sigma_{\rm stat}^2}.
\end{equation}
We assumed a normal distribution with a standard deviation of $\sigma_{\rm stat}$ here as most uncertainties in our preliminary modellings suggested it to be a valid approximation. If there is also a systematic component in uncertainties, the real likelihood difference should be
\begin{equation}
    \Delta \log \mathcal L' = \frac12\frac{x\Delta x + \Delta x^2}{\sigma_{\rm stat}^2+\sigma_{\rm sys}^2},
\end{equation}
where also a systematic uncertainty of the modelling $\sigma_{\rm sys}$ is added. We may write
\begin{equation}
    \Delta \log L' = C\cdot \Delta \log L \label{eq:C_def},
\end{equation}
where $C$ is the correction to include systematic uncertainty with the value:
\begin{equation}
    C = \frac{\sigma_{\rm stat}^2}{\sigma_{\rm stat}^2+\sigma_{\rm sys}^2} = \left( 1 + \frac{\sigma_{\rm sys}^2}{\sigma_{\rm stat}^2} \right)^{-1}.
\end{equation}
The advantage of including the systematic uncertainties in likelihood calculations is the ability to model all systematic uncertainties simultaneously and to preserve more accurate posterior distribution than adding systematic component. 

Numerical values of $C$ are calculated based on comparing modelling results and corresponding actual values (see Sect.~\ref{sec:mock_conclusion}).

\subsection{ Selection of acceleration field}
To apply the method to the data, we must first assume a sufficiently flexible form of the acceleration field. As the sizes of the studied regions are small (their radii are about $1$ kpc) the acceleration field can be described in a very concise way. We chose a Taylor expansion like forms\footnote{As most functions can be expanded to the Taylor series, comparison with global MW models can be done simply by expanding the global MW acceleration field into a Taylor series around the SN and by comparing thereafter expansion coefficients with our fitted values.}: average components of the field with added correction terms in directions where acceleration components change significantly
\begin{eqnarray}
    a_R &=& A_R + A_{R,R}\cdot\Delta R  \label{eq:accfield1} \\
    a_\theta &=& A_\theta\\
    a_z &=& A_z + A_{z,z}\cdot z . \label{eq:accfield3}
\end{eqnarray}  
Here we used usual cylindrical coordinates, 
\begin{equation}
    \Delta R = R - R_{\rm cnt} , \label{eq:accfield4}
\end{equation}
and $R_{\rm cnt}$ corresponds to the centres of regions. We have denoted $(a_R, a_\theta, a_z)$ as the acceleration vector components. Capital letters $A_R, A_\theta, A_z$ are average acceleration components within a region (zero terms of expansion), and $A_{R,R}, A_{z,z}$ characterize the radial and vertical changes with radius and height above the plane. The acceleration field (\ref{eq:accfield1}) -- (\ref{eq:accfield3}) is irrotational, hence eligible to describe the fields generated by gravity. 

\begin{figure*}
    \centering
    \includegraphics[width=\textwidth]{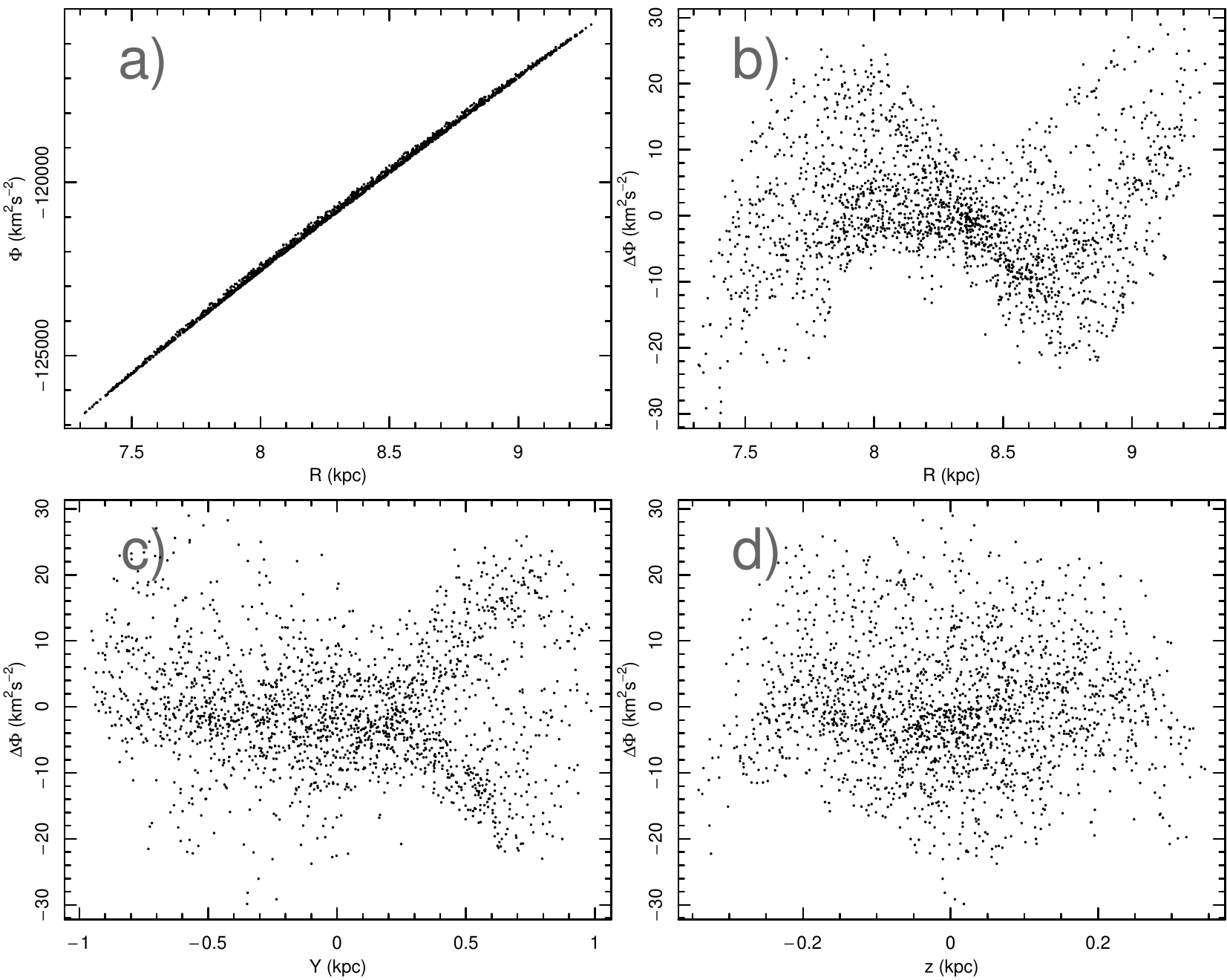}
    \caption{The adequacy of the used functional form of the acceleration field to describe potential. To avoid cluttering the figure, we plotted a subsample of $2000$ points out of $\sim 60\,000$. Panel a) shows the true potential as a function of radius; the narrowness of the correlation shows that just this dependence is the dominant source of potential variance. Panels b-d show the residuals of the true  and best-fitted potential $\Delta \Phi = \Phi - \Phi_{\rm pred}$. In panel b) we show the radial dependence; panel c) shows tangential dependence with $Y$ defined as $R\Delta\varphi$; and panel d) describes residuals in the vertical direction. }
    \label{fig:funform_adequacy}
\end{figure*}

Acceleration field in form of (\ref{eq:accfield1}) -- (\ref{eq:accfield3}) corresponds to the potential
\begin{equation}
\Phi = \Phi_0 + A_{R0}R + \frac12A_{R, R}(\Delta R)^2 + A_{\theta}R\Delta\theta + A_{z}z + \frac12A_{z, z}z^2.
\end{equation}
We made the adequacy test for the functional form based on SR1 that contained $63\,646$ stars. We determined are the linear terms in expansion (\ref{eq:accfield1}) -- (\ref{eq:accfield3}) sufficient to describe adequately  potential changes in simulation data. We compared values taken from simulation $(\Phi)$ and calculated from the best-fitted test model $(\Phi_{\rm pred})$. 
Results of the test are shown in Figure \ref{fig:funform_adequacy}. 
Panel a) shows the range of the true gravitational potential values $\Phi$, which cover $11353\,{\rm km^2s^{-2}}$. Panels b)-d) describe the residuals of the potential fit $\Delta \Phi = \Phi - \Phi_{\rm pred}$. As the residuals cover the range of only $63\,{\rm km^2s^{-2}}$ we may conclude that our used functional form describes gravitational field with an accuracy of $0.6\%$. We point out that the simulation did not contain a gas disc that may amplify the importance of vertical acceleration changes.

\subsection{Modelling in a rotating reference frame}\label{sec:ref_frame_modelling}

There is a technique to convert non-stationary modelling into stationary by modelling the stellar system in a rotating reference frame e.g. \citep{Jung:2015}. Let us denote the angular speed of the reference frame as $\Omega_{\rm fit}$. In case $\Omega_{\rm fit}=\Omega_{\rm p}$, and a single pattern speed is sufficient to describe all the non-stationarity, then the exercise reduces to a stationary one. Although this technique is efficient in describing bar-induced non-axisymmetry, we are using it for another purpose.

When the star moves in the studied region, it spends there about $\approx 3\,{\rm Myr}$. In case the acceleration is $\approx6000\,{\rm km^2s^{-2}kpc^{-1}}$, the change of stars velocity is $18\,{\rm km\,s^{-1}}$. For studying more subtle effects from acceleration (e.g. the acceleration caused by the bar), the changes of velocity are about $3\,{\rm Myr}\times 200\,{\rm km^2s^{-2}kpc^{-1}}\approx 0.6 \,{\rm km\,s^{-1}}$. Detecting distribution shifts of this small is troublesome. In case we are modelling the system in a rotating reference frame, stars take longer to pass the studied region, the subtle acceleration effects increase, and we can make more robust inferences. 

When modelling in the rotating reference frame with angular velocities $\Omega_{\rm fit}$, then the time average star spends in the studied region increases by a factor $\frac{\Omega_{\rm star}}{\Omega_{\rm star}-\Omega_{\rm fit}}$, where we have denoted the angular speed of an average star with $\Omega_{\rm star}$. The changes of acceleration from a component moving with $\Omega_{\rm p}$ would decrease by the factor of $\frac{\Omega_{\rm p}-\Omega_{\rm fit}}{\Omega_{\rm p}}$. In case the angular speed of the asymmetry and the angular speed of the star are similar, then the overall effect from time-dependence of the potential remains the same. In subsequent sections, we provide modelling results and their dependence on the rotation of the reference frame. 

\subsection{Selection of the non-stationarity description}
The novelty of the present analysis is an introduction of a non-stationarity in the orbital arc method. More specifically, while calculating small orbital arcs in a selected region and thereafter calculating probabilities to find a star in these orbits we introduced the term $ht$ in Eq.~\eqref{eq:dP_const_used}. For every orbit the constant $h$ has a certain value. In a selected region there are a multitude of orbits and probably many different values of $h$. For example, we cannot assume that halo and disc are similarly perturbed at SN; hence we must implement a possibility that certain orbital parameters may determine $h$ values. Thus, although $h$ is assumed to be a constant for a given orbit, while looking to the whole region $h$ is a function of orbits. We assume that $h$ is a smooth function of integrals of motion, particularly the angular momentum $L_z$ integral. The choice of angular momentum as the dependent variable is motivated by the results of \citet{Antoja:2018} where they show substructure comes forth by analysing rotational velocities (in case of small regions, rotational velocity and angular momentum $L_z$ have almost one-to-one relation). The functional form for $h$ we selected in a way that most stars are populating their orbits stationary, but some with a certain angular momenta have elevated to non-stationarities:
\begin{equation}
    1 + ht  = 1 + h(L_z)t = 1 + h_{Lz}\exp{\left[-\frac{(L_z - L_{z0})^2}{L_{z\sigma}^2}\right]t}.\label{eq:Lz_pert_form}
\end{equation}
The $L_{z0}$ describes the angular momentum value around which orbits are populated most non-stationarily. The $\sigma_{Lz}$ shows the spread of the non-equilibrium orbits, the $h_{Lz}$ shows the maximum level of non-stationarity of orbits. 

\subsection{Implementation of mock modelling and conclusions}
To test the limits of the method, we modelled the mock data repeatedly with parameters emphasizing different aspects of the model. These aspects include the rotation of the reference frame (selection of $\Omega_{\rm fit}$), description of non-stationarity (selection of $h$) and inclusion of systematic effects (selection of $C$).

First two aspects are inter-related. The average angular speed of the region is $\Omega_{\rm star}\approx 25\,{\rm km\,s^{-1}kpc^{-1}}$. To increase the time stars spend in the region the $\Omega_{\rm fit}$ should be as close to $25\,{\rm km\,s^{-1}kpc^{-1}}$ as possible. But let us look closer. If $\Omega_{\rm fit}=0$ all the stars in the disc region enter the region from one side and exit from the other. Once $\Omega_{\rm fit}$ approaches $\Omega_{\rm star}$, the stars exit the studied region in a random direction. If $\Omega_{\rm fit}>\Omega_{\rm star}$ stars move in the region in reverse compared to the region that stands still. But now issue arises in the interpretation of $h$ in the modelling. We wish to keep the interpretation $h \approx (\partial f/\partial t)/f$, but this relation does not hold well if $\Omega_{\rm fit}\ne 0$ as the orbit extrapolation "meet" different sets of stars for $\Omega_{\rm fit}$ values. Only in case the $h$ is a very smooth function of integrals of motion, the relation still holds.

Thus, we selected priors for the parameters in analytical form of $h$ (Eq.~\eqref{eq:Lz_pert_form}) accordingly. The $L_{z0}$ had to be within the $10$ and $90$ percentiles of the stellar angular momenta so that in case of zero perturbations, there would be no degenerate solutions in ranges where there are no stars. The width of the distribution for angular momenta of the non-equilibrium was between $L_{z\sigma}$ $100$ and $700\,{\rm km\,s^{-1} kpc}$, and the $h_z$ values was between $\pm50\,{\rm Gyr^{-1}}$. During the mock data modelling, we did not reach these prior limits.

\subsubsection{Conclusions from the acceleration modelling}
\label{sec:mock_conclusion}
\begin{figure*}
    \centering
    \includegraphics{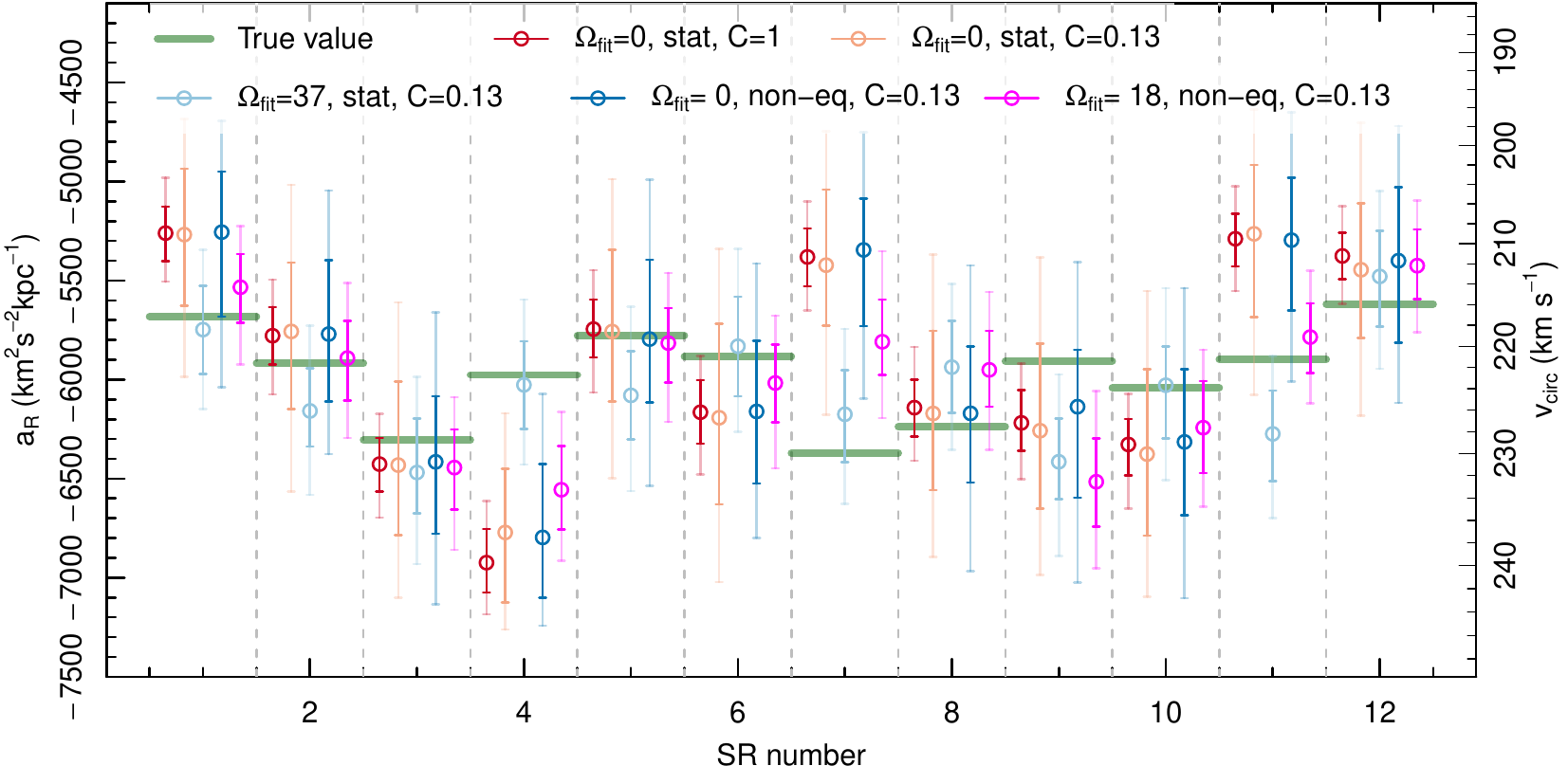}
    \caption{The recovery of radial acceleration $a_R$ from modelling simulated data. For each region, we present a set of results from different modellings with $1\sigma$ and $2\sigma$ uncertainties. For each region, from the left, corresponding points are (1) nonrotating reference frame in stationary modelling to estimate systematic uncertainties; (2) the test how well the inclusion of systematic uncertainties works; (3) modelling in case of the reference frame corresponding to the patter speed; (4) non-stationary modelling where the relation $h\approx (\partial f/\partial t)/f$ holds well; (5) modelling with the best fit parameter set. }
    \label{fig:mockfit_ar}
\end{figure*}

\begin{figure*}
    \centering
    \includegraphics{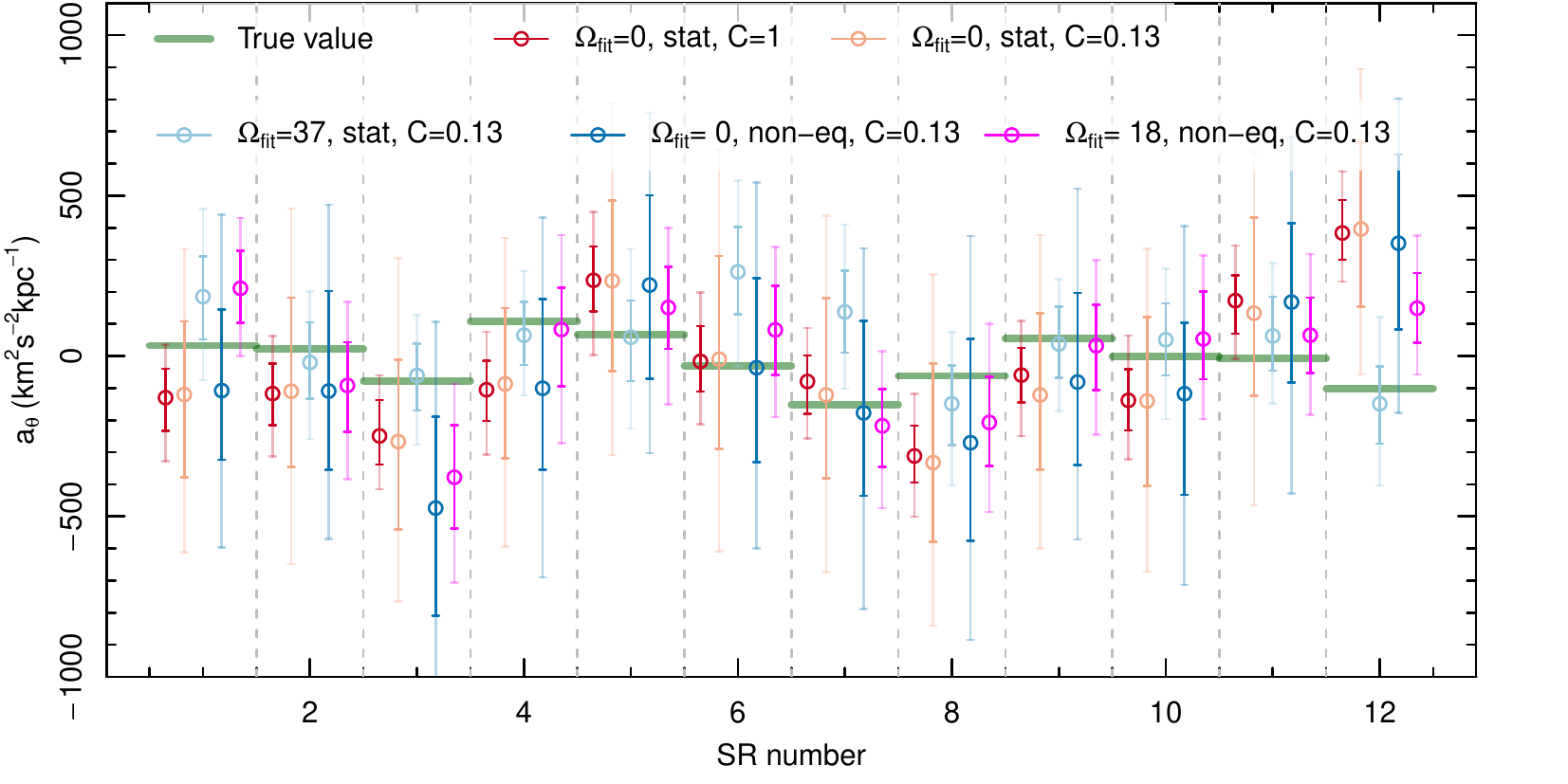}
    \caption{Same as Fig.~\ref{fig:mockfit_ar}, but calculated based on recovery of $a_\theta$.}
    \label{fig:mockfit_at}
\end{figure*}

We made five different fittings of mock data. In each case we had $12$ regions. The results are given in Figs.~\ref{fig:mockfit_ar} and \ref{fig:mockfit_at}. Overall, it seems that calculation of $a_R$ is more biased and is more difficult to fit, hence we mostly emphasize our analysis based on $a_R$. The analysis based on $a_\theta$ is very similar. 


First, we modelled SRs in a non-rotating reference frame ($\Omega_{\rm fit} = 0\,{\rm km\,s^{-1}kpc^{-1}}$) without considering systematic errors ($C=1$) and assuming stationarity ($h=0$). These results are depicted as red circles in Figs.~\ref{fig:mockfit_ar} and \ref{fig:mockfit_at}. One can see that in four cases (2, 3, 5, 8), the fit is quite good, and the model and actual value differences are within $1\sigma$ uncertainties. In addition in three cases (6, 9, 12) differences are within $2\sigma$ uncertainties. In regions 1, 4, 7, 11 the uncertainty is clearly under-estimated. Let us look at the positions where the regions situate from Fig.~\ref{fig:mock_illustration}. In a strongly barred galaxy, we would expect that regions that are $180^\circ$ shifted should give a mismatch that matches the initial ones exactly. This matching will only happen if the bar or its induced non-equilibrium is the cause of the offset. In our case, the region pairs that are $180^\circ$ off are $1-7$, $2-8$ etc. For the pair $4-10$, region 4 has a high mismatch, while the other is $\approx 2\sigma$ offset. For the $1-7$ pair, also one offset is significantly higher than the other, for the pair $11-5$, the offsets also do not match at all. As these offsets do not match, we conclude that it is more likely that the origin of these offsets is not physics related but noise and sampling related. 

Next, based on the non-rotating stationary modelling, we estimated the level of systematic uncertainties that statistical uncertainties do not cover. We applied the recipe described in Sec.~\ref{sec:mock_sys_uncert} to estimate that likelihood ratios should be multiplied by a certain constant $C$. The numeric values of $C$ are calculated based on comparing modelling results without systematic uncertainties ($C=1$, red circles) and their distances to the actual values (green lines). Using $C$ as a free parameter we selected $C$ values that cover the offset best. In Table~\ref{tab:C_value}, we present the $C$ value calculated based on acceleration component differences from actual value $A - A'$ and $\sigma_{\rm stat}$ values to fit the expectation $\mathbb{E} (A - A')^2 = \sigma_{\rm stat}^2 + \sigma_{\rm sys}^2$. Since the values differ for different parameters, we select the value in between of most important parameters we would like to recover precisely: $A_R$ and $A_\theta$. Then we adopted the value $C = 0.13$ and calculated accelerations again. Corresponding accelerations are depicted by orange circles. 

\begin{table}
    \centering
    \caption{The $C$ values calculated based on different acceleration components. Values over $1$ indicate that the statistical uncertainty was overestimated from the modelling. }
    \label{tab:C_value}
    \begin{tabular}{l|ccccc}
    Variable & $A_R$ & $A_{z,z}$ & $A_\theta$ &  $A_{R,R}$ & $A_z$ \\
    \hline
     $C$ &  0.088  &  2.26  &  0.20 &  0.57&  3.3 \\
     \hline
    \end{tabular}
\end{table}

Next, we test the influences of the speed of the reference frame. In the figures, we show the modelling with condition $\Omega_{\rm fit}=\Omega_{p} = 37\,{\rm km\,s^{-1}kpc^{-1}} $ with light blue circles and results are clearly better for the regions that had previously substantial offset from the actual value (see especially regions 1, 4, 6, 7 and 10). The only exceptions are regions 5, 8, 9. 

For the next test, we add the non-equilibrium possibility to the modelling ($h\ne 0.$).  We depict the non-equilibrium models as dark blue circles in the figures. Comparing the results with the corresponding stationary version (orange circles), we do not see significant changes. We conclude that in the $R\approx8\,{\rm kpc}$ regions, the non-equilibrium has less significance than effects from acceleration. An alternative explanation is that the number density of stars is too little to model in such depth that the non-equilibrium starts to dominate over acceleration components in a small region. The third explanation is that the functional form of $h$ is not sufficiently diverse to describe actual non-equilibrium. In the next section, we will dissect the recovered non-equilibrium values more thoroughly.

In the last test-run, we adapted $\Omega_{\rm fit}=18\,{\rm km\,s^{-1}kpc^{-1}}$ to maximise the time star spends in the studied region, included non-stationarity and used the $C=0.13$ value (magenta circles). We could recover the actual values well in nine cases. In the case of the three problematic regions (4, 7, 11) the $\Omega_{\rm fit}=\Omega_{p}$ gives better results in two cases, while in SR 11, the $\Omega_{\rm fit}=18\,{\rm km\,s^{-1}kpc^{-1}}$ is better. Overall, we conclude that we can adequately recover acceleration in most cases, and if the mismatch is caused by sampling noise, we do not expect similar problems in the observational application.

\subsubsection{Conclusions from the non-equilibrium modelling}
\label{sec:mock_h_modelling}
We included the non-equilibrium term $h$ for $\Omega_{\rm fit} = 18$ and for $\Omega_{\rm fit} = 0$. In both cases, we calculated a set of parameters that describe the systems underlying non-equilibrium. Before discussing derived results we would like to note what to expect when modelling non-equilibrium in a rotating reference frame.

Let us imagine a density gradient caused by non-equilibrium. Densities are proportional to time (oPDF) and time in a studied region is determined by the region's size and the velocities. As we use at present angular velocities let us denote the $\Delta \varphi$ as the angular size of the studied region. Then for an average star, it would take $t_0 = \Delta\varphi / \Omega_{\rm star}$ to leave the region. In a rotating reference frame, the time would be $t_{r} = \Delta\varphi / (\Omega_{\rm star}-\Omega_{\rm fit})$, indicating that times and hence also density gradient would be increased by a factor of $\approx \Omega_{\rm star}/(\Omega_{\rm star}-\Omega_{\rm fit})$. As times needed to produce a gradient would change by this factor, the $h$ values should also be modified by the same amount. These considerations are not rigorous, nor are they intended to be, as not all motions in the disc are rotation dominated, especially in a rotating reference frame where $\Omega_{\rm fit}\rightarrow\Omega_{\rm star}$.   In case of $\Omega_{\rm fit}=18$ that is used in simulations the factor $\Omega_{\rm star}/(\Omega_{\rm star}-\Omega_{\rm fit})$ is  $\approx 3.6$. 

\begin{figure}
    \centering
    \includegraphics{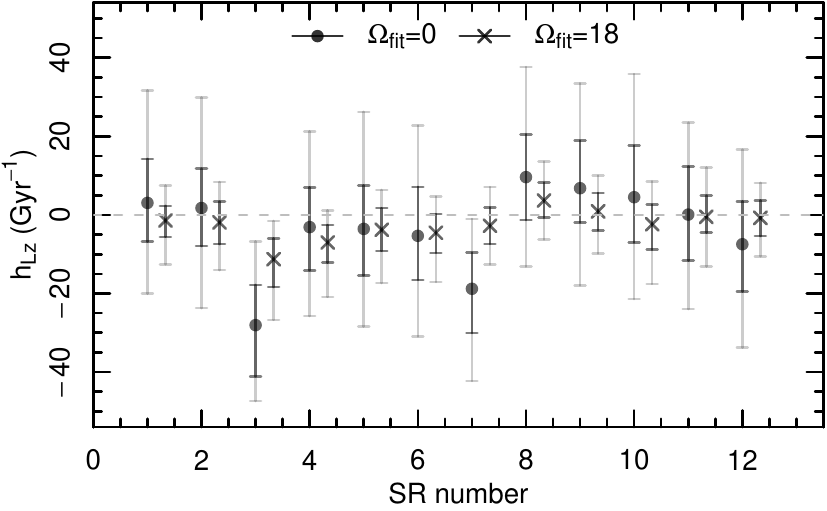}
    \caption{The fitted $h_{Lz}$ parameter values and uncertainties for modelling in both non-rotating and rotating reference frames. One can notice a bias toward lower absolute values, which is expected to be $\approx3.6$ times compared the rotating and non-rotating frames (see main text). }
    \label{fig:hLz_values}
\end{figure}
In Fig.~\ref{fig:hLz_values} we provide the amplitude values of the function $h(L_z)$ (see Eq.~\eqref{eq:Lz_pert_form}) for $\Omega_{\rm fit} = 0$ and $18$. We see that the $\Omega_{\rm fit}=18$ based values are closer to zero line than the $\Omega_{\rm fit}=0$ based values. Based on the considerations from the previous paragraph, this is expected. But we see also that most of the values are in fact zeros within the error. This means that in the simulation, the regions at $R=8.3\,{\rm kpc}$ are not sufficiently perturbed to recover the perturbation from such a small region. Thus, the levels of non-equilibrium are smaller than $|(\partial f/\partial t)f^{-1} |\lesssim 0.03\,{\rm Myr^{-1}}$ or three per cent per Myr. 

\begin{figure}
    \centering
    \includegraphics{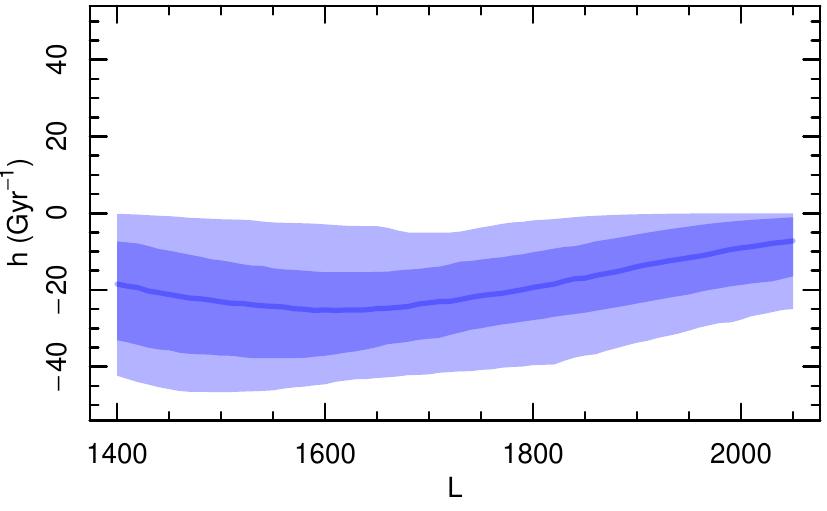}
    \caption{Distribution of $h$ as a function of angular momentum $L_z$ for the mock SR3 region.  The blue colors describe the median value, $\pm1\sigma$ and $\pm2\sigma$ regions. These regions are calculated by constructing single lines of posterior parameter samples and selecting their quantile lines Hence, they do not have to follow the Eq.~\eqref{eq:Lz_pert_form}. }
    \label{fig:SR3_h_fun}
\end{figure}
The $h_{Lz}$ is just one parameter describing the non-equilibrium function $h$. The entire function is shown in Fig.~\ref{fig:SR3_h_fun} in case of the region SR3 (non-rotating frame). We picked this region as it had a large $h_{Lz}$ value. The median and uncertainty corridor is constructed by calculating all the $h(L_z)$ function lines from the MCMC posterior points and estimating quantiles of each $L_z$ value. It is seen that $h$ values are negative in this region. 

\begin{figure}
    \centering
    \includegraphics{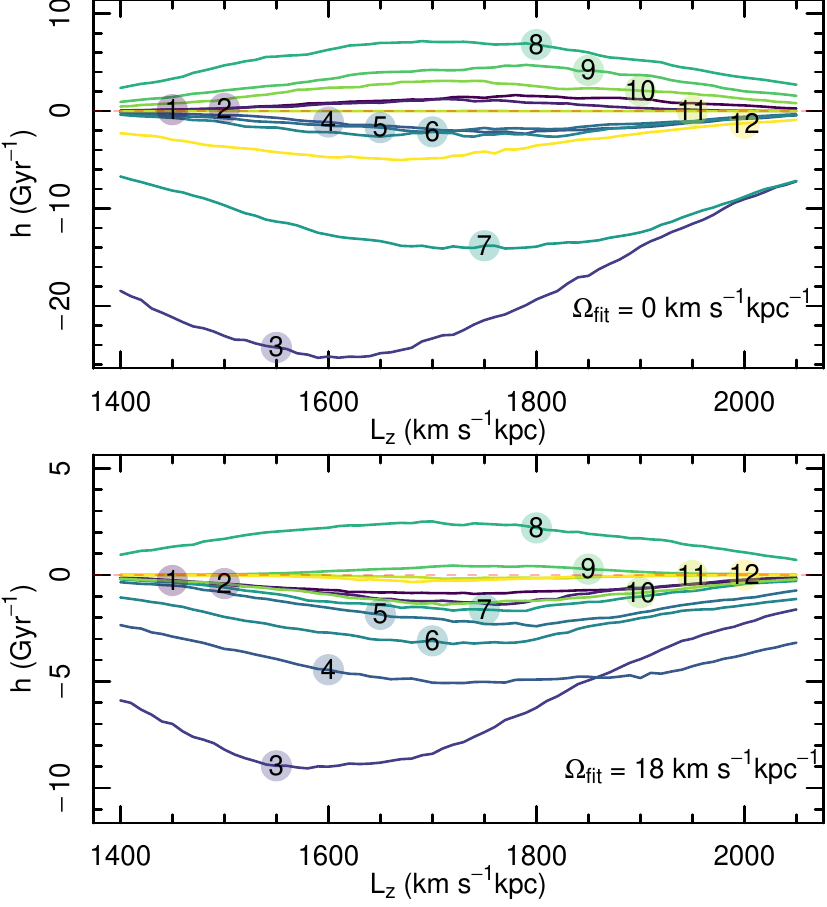}
    \caption{The medians of the function $h$ for all regions as a function of angular momentum $L_z$. Region numbers are in the circles on top of each line. The top panel shows results for modelling in case of $\Omega_{\rm fit}=0$, and the bottom one for $\Omega_{\rm fit}=18\,{\rm km\,s^{-1}kpc^{-1}}$. The differences of $\sim3.6$ times are expected due to the rotation of the reference frame. } 
    \label{fig:h_amplitudes_mock}
\end{figure}
In Figure \ref{fig:h_amplitudes_mock} we show the median lines of all the regions, not just one as in Fig.~\ref{fig:SR3_h_fun}. Again, we see that $h$ values depend strongly on the $\Omega_{\rm fit}$ values as described earlier in this section. The ones that are modelled with $\Omega_{\rm fit}=0$ are the ones that have the interpretation $h\approx (\partial f/\partial t)f^{-1}$. We see from the top panel, that $h$ values make a decreasing trend between SR8 and SR12. A similar trend is seen also between SR1 to SR6 (except for SR3). It seems that over half of the galaxy, the phase space density grows due to non-equilibrium and over the other half, it decreases (see Fig.~\ref{fig:mock_illustration}). This trend is similar to the $m=1$ asymmetry, but for $\partial f/\partial t$. The $\Omega_{\rm fit}=18$ modelling semi-confirms it, but one shoud keep in mind that 1:1 relation between $\Omega_{\rm fit}=18$ and $\Omega_{\rm fit}=0$ holds only in case of zero-dispersion dynamics, which is not the case. Due to weak statistical significance, we are not able to make any firm conclusions about it.

\section{Non-stationarity of the Milky Way in the Solar neighbourhood from Gaia DR2 data}\label{sec:gaia_fit}

In \citet{Kipper:2020} we used the orbital arc method to calculate the acceleration field in the Solar Neighbourhood. The model was a stationary one. In present paper (see Sect. 2) we improved the method to include also possible non-stationarities of the galaxy. Here we re-model the SN region to find out how perturbed region we are living in. Test of the method with mock data presented in Sect.~3 confirmed us that the method allows us to conclude that the method is reliable.

For the Solar velocity we use the latest values that are based on Gaia EDR3 reflex motion minimisation \citep{Malhan:2020}; the values are $V_{\odot, R} = 8.88, \quad V_{\odot, \theta} = 241.91, \quad V_{\odot, z} = 3.08\,{\rm km\,s^{-1}}$.

\subsection{Implementation of the modelling with Gaia data}

We use a similar dataset as in \citet{Kipper:2020}: the region is a flux-limited (in $J$-band  up to $10.25^m$) oblate spheroid, with semi-height of $0.4$~kpc and radius $0.494$~kpc containing $525\,264$ stars from Gaia DR2 \citep{gaia1, gaia2} with StarHorse \citep{starhorse} distance estimations. 
The size of the region determines the time-scale for how long the non-stationarity estimates can be handled as valid and be extrapolated. The average velocity of stars in the region is about $\approx 240\,{\rm km\,s^{-1}}$ and the diameter of the region is $\approx 1\,{\rm kpc}$, causing stars to be in the region for $\lesssim 4\,{\rm Myr}$. This is the maximum time interval the derived  estimates for the non-equilibrium parameter $h$ should be used. 

To use the orbital arc method, one needs as an input all six phase space coordinates of stars. As the Gaia observations provide at present radial velocities for only a moderate number of stars it observes, this causes a selection bias. For stars with $m_g \simeq 4^m$ about $50\%$ of stars have a measured radial velocity, while for stars with $\simeq 11^m$ the completeness is $\sim80\%$ (see \citet{Katz:2019}). In addition, there is a significant direction-dependence. In the previous paper \citep{Kipper:2020} we used a match with 2MASS magnitudes that are not attenuation sensitive and constructed dataset for modelling is based on these. 

\begin{figure}
    \centering
    \includegraphics{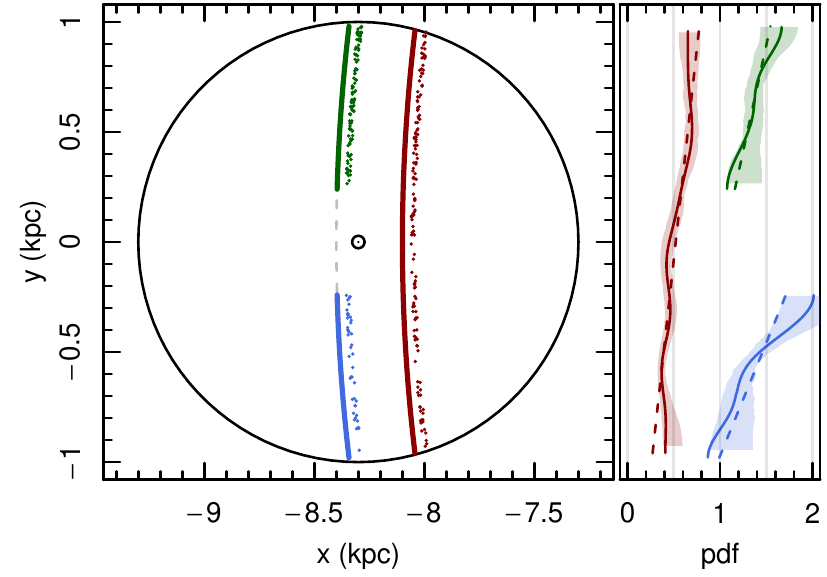}
    \caption{An illustration of a side-effect of the bright-end sample cut. The left-hand panel shows how the brightness limit splits an intact orbit (the red one) into two (green and blue ones) by getting too close to the Sun so that apparent magnitude increases. The stars are populated along the orbit (but shifted for illustration purposes) to mimic the non-equilibrium density population. The right-hand panel shows the probability density distributions (PDF) along the $y-$direction, dashed line as the slope data was generated with. Differences of slopes are apparent due to not normalisation to $1$. One can notice higher uncertainties in case the orbit is split (the wavy sampling noise do not cancel out sufficiently), causing an overall hindrance in modelling.  }
    \label{fig:illustration_brightend}
\end{figure}

In the previous paper, we adapted a flux-limited approach by choosing a superposition of volume-limited samples. We added a similar cut to cope with the bright-end selection: we removed all the stars that are fainter than some bright-end $J$ band magnitude limit ($J_{BE}$). The validity of this procedure is the same as when cutting the dim-end of the sample, and the reasoning is more thoroughly covered in \citet{Kipper:2020}. Removal of the bright end is done in the same way in modelling: the orbit is integrated until the star exits the studied region or its brightness is not between the used magnitude limits. Although it provides the same overall results, the accuracy of the model is more diminutive. As an illustration we present in Fig.~\ref{fig:illustration_brightend} two (oversimplified) cases: one where an orbit runs through the region intact and the case where the brightness limit "cuts" the orbit into two. In the right-hand panel, we provide the density distributions and the corresponding uncertainty corridor for both cases, along with the actual slope we generated the data. As the split orbital PDF distribution has higher uncertainties than the unsplit one, the overall modelling through these is more uncertain, and the overall accuracy of the model is reduced. We used two different limits $J_{\rm BE}$: $9^m$ (default) and $7^m$ (denoted as less-conservative). The former sample contains $352\,656$ stars and latter one $481\,464$ stars.

As the selection function has a significant impact on data distribution, we added a supplementary way to cope with it. For the orbit calculations, we multiplied the $h$ with a term $h_{\rm selection}$, which is an inverse of the fraction of stars observed compared to completeness, hence after division, we get the incompleteness corrected orbital populations. Based on the \citet{Katz:2019} results, we approximated it to be a linear function of apparent magnitude $h_{\rm selection} = h_{s0}+h_{s1}m$. Hence, a stellar orbit is populated with the density 
\begin{equation}
 \rho\propto {\rm d}t(1+h)(h_{s0}+h_{s1}m),
\end{equation}
where $m$ is the apparent magnitude of that orbital point\footnote{apparent magnitude is determined based on Solar location, and the absolute magnitude of the star that's orbit is being calculated.}. Again, due to the normalisation of PDF of each orbit, we selected $h_{s0}$ to be $1$. The parameter describing the selection ($h_{s1}$) is a free parameter in modelling. We expect it not to be confusable with physical non-equilibrium as it influences orbital densities with spherical symmetry around the Sun, compared to along the rotation for non-equilibrium.

The modelling is done in a similar manner as we tested the method with mock data \citep{Kipper:2019} although the statistical inference at present is done with a finer grid using $30^2=900$ grid cells $G=32$ times. We set a limit that no cell should have less than $10$ stars to avoid too small grid cells, and high relative Poisson noise. The likelihood is maximised using \textit{Multinest} \citep{MN1, MN2, MN3} code with $100-500$ live points. We modelled the region with parameter $C = 0.13$ from Eq.~\eqref{eq:C_def}, which might give overestimated uncertainties in case the cause of uncertainties is sampling noise. 

As the SN region is quite far from the bar, the acceleration field should be sufficiently well described with a simple analytical form. Hence, we select the form \eqref{eq:accfield1}-\eqref{eq:accfield3} to describe the acceleration field. For the non-equilibrium term, we used the same form as in simulation data Eq.~\eqref{eq:Lz_pert_form}.

We used three different reference frames when modelling: $\Omega_{\rm fit} = \{0, 20, 40 \}\opunit$. These values provide results where we emphasize the recovery of non-equilibrium parameters or acceleration parameters. The values differ slightly from the ones we used in mock analysis since the velocities of MW are slightly larger. We chose the $\Omega_{\rm fit}=20\opunit$ since MW stellar angular velocity is $\approx 29\opunit$ and based on the mock results we chose the $\Omega_{\rm fit}$ to be less than this value, but still large enough to increase the time stars spend in the region in a rotating reference frame.  

\subsection{The non-stationarities in the Solar Neighbourhood}\label{sec:gaia_results}
\begin{table*}
    \caption{Calculated accelerations and perturbation levels in Solar Neighnourhood. In the first two columns assumed angular velocity of the reference frame and limiting $J$-band magnitude is given, in subsequent five columns acceleration field coefficients are given, in the last column perturbation amplitude parameter $h_{Lz}$ is given. }
    \label{tab:obsresults}
    \centering
    \begin{tabular}{ccccccc|c}
    \hline
    \hline
$\Omega_{\rm fit}$ & $J_{\rm BE}$ & $A_{R}$ & $A_\theta$ & $A_z$ & $A_{R,R}$ & $A_{z,z}$ & $h_{Lz}$  \\
         ${\rm km\,s^{-1}kpc^{-1}}$ & $m$ & $\accunit$ & $\accunit$ & $\accunit$ & $\accunitperkpc$ & $\accunitperkpc$ & ${\rm Gyr^{-1}}$ \\
         \hline
$ 0$ & $9$ & $-6502^{+605}_{-539}$ & $-304^{+331}_{-360}$ & $223^{+169}_{-237}$ & $494^{+1670}_{-1544}$ & $-3012^{+1671}_{-1709}$ & $-13^{+49}_{-43} $\\
$ 0$ & $7$ & $-6462^{+273}_{-238}$ & $-348^{+196}_{-200}$ & $191^{+129}_{-121}$ & $262^{+1237}_{-1333}$ & $-2789^{+1196}_{-1193}$ & $79^{+59}_{-52} $\\
$ 20$ & $9$ & $-6354^{+198}_{-198}$ & $-245^{+124}_{-118}$ & $106^{+94}_{-102}$ & $-237^{+1065}_{-899}$ & $-2888^{+714}_{-788}$ & $-7^{+17}_{-22} $\\
$ 40$ & $9$ & $-6253^{+231}_{-237}$ & $119^{+178}_{-219}$ & $-93^{+136}_{-135}$ & $-895^{+1001}_{-745}$ & $-3073^{+954}_{-976}$ & $-15^{+22}_{-43} $\\

\hline
    \end{tabular}
\end{table*}
\begin{figure}
    \centering
    \includegraphics{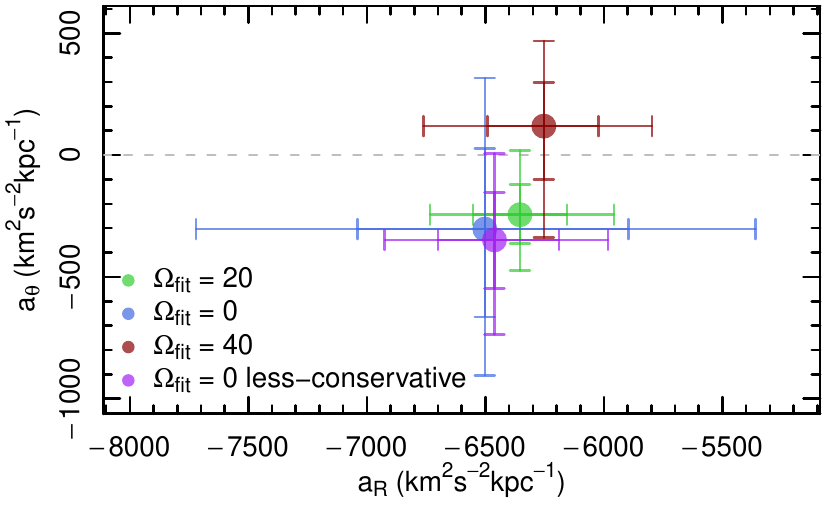}
    \caption{The two main components of recovered acceleration field. The horizontal axis show radial, and vertical axis tangential acceleration components. Different color lines show modelling results in different reference frames or selection limits. We provide both $1\sigma$ and $2\sigma$ uncertainties. }
    \label{fig:obs_acc}
\end{figure}
\begin{figure}
    \centering
    \includegraphics{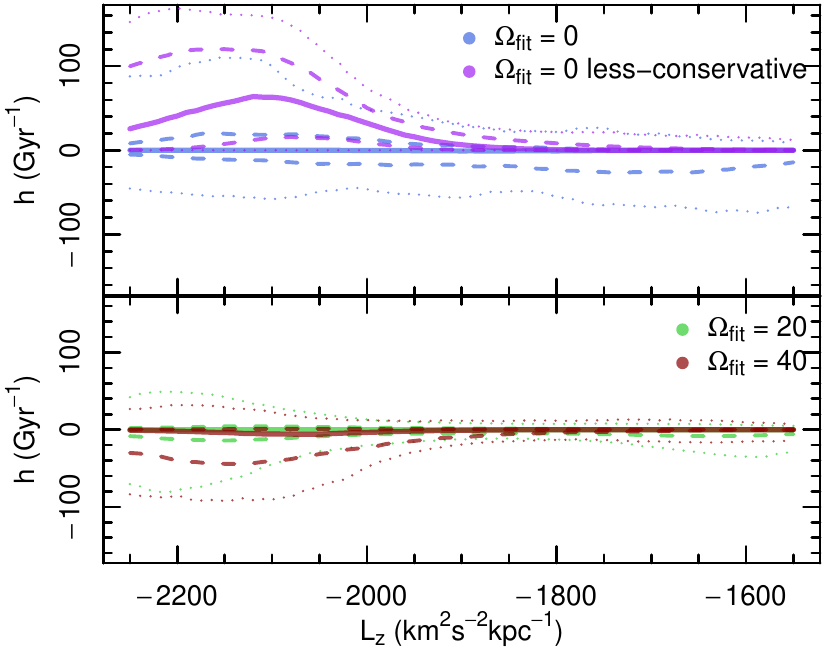}
    \caption{The levels of calculated non-equilibrium function $h(L_z)$ in SN. Dotted thin lines show $\pm2\sigma$, dashed ones $\pm1\sigma$ and continuous line give the median values similar to Fig.~\ref{fig:SR3_h_fun}. The top panel show the results where $h\approx \partial f\partial t$ interpretation holds, the bottom ones where it doesn't (and is even reverse, see Sect.~\ref{sec:mock_h_modelling})}
    \label{fig:obs_h}
\end{figure}
Modelling the SN by including non-stationarities, we can update the values of acceleration and assess the importance of non-stationarities. 
We show the calculated acceleration field in Table~\ref{tab:obsresults} and the main components of it in Fig.~\ref{fig:obs_acc}.

For the radial acceleration $A_R$ we found the weighted mean of all the radial accelerations in Table~\ref{tab:obsresults} to be $\langle A_R \rangle = -6360\accunit$. %\pm126
All different implementations of the modelling (mostly the angular speed of the reference frame $\Omega_{\rm fit}$) give consistent results. Derived $\langle A_R \rangle$ corresponds to the circular velocity of $228.4\,{\rm km\,s^{-1}}$, which is slightly less than the usual ones from the literature (e.g. $v_{\rm c} = 238 \pm 9 {\rm km\,s^{-1}}$ from \citet{Schonrich:2012}). The main reason is the difference of the used Solar motion, \citet{Malhan:2020} compared to \citet{Schonrich:2012}, being $10.33\,{\rm km\,s^{-1}}$. Taking this into account the circular velocities are consistent with each other. The radial acceleration derived in \citet{Kipper:2020} was $-6214\accunit$. The main difference between these results is the inclusion of non-equilibrium. We can conclude that the equilibrium assumption gives in the SN case the difference by $\sim2\%$ for radial acceleration. 

The average tangential acceleration $\langle A_\theta\rangle$ is $-201\accunit$. Contrary to the radial accelerations, the tangential ones varied clearly more between different modelling implementations. In order to find out which implementation is a preferred one, let us remind our tests with mock data. In Fig.~\ref{fig:mockfit_at} we provide the recovery precision for different modelling settings. Comparing the cases where $\Omega_{\rm fit}$ is $0\opunit$, $18\opunit$ or $37\opunit$, the results are mostly similar, but still in some cases differ. The differences are largest for cases where the rotation of the frame matches the bar pattern speed: mostly $\Omega_{\rm fit}=\Omega_{\rm bar}$ has a similar, better or significantly better fit than others; however in two cases (regions SR 6 and 7) fits are worse or significantly worse. Based on the mock analysis, we saw that the average consistency is improved when $\Omega_{\rm fit}$ is close to stellar velocity. Taking this into account, we favour the results with $\Omega_{\rm fit} = 20\opunit$, i.e. $A_\theta = -245^{+124}_{-118}\accunit$. If the torque from the bar causes this acceleration, then this acceleration corresponds to the mass of the bar of $1.3\times10^{10}\,{\rm M_\odot}$ (see details from \citet{Kipper:2020} with the assumed bar density profile from \citet{Wegg:2015}).

Let us introduce the parameter $h_{s1}$ describing a mismatch due to apparent magnitude cut caused by the selection function. The numeric value we model for the $h_{s1}$ is quite uncertain but has an average $\approx -0.01$. In modelling, the value does not drop below $-0.025$ but can be as high as $0.1$ in the posterior sample. For the interpretation, the average numeric value $h_{s1}$ the bright end and dim end selection (extreme values) gave number density differences of $1.5\%$ or $3.5\%$, depending on the used apparent magnitude limit for selection. From the vast range of acceptable values to the modelling and no correlation between acceleration parameters and $h_{s1}$, we conclude that the selection function did not contribute significantly to the modelling. 

While describing the mock data modeling (Sect.~\ref{sec:ref_frame_modelling}) we mentioned that when handling $\Omega_{\rm fit}$ and function $h$ one needs to distinguish two separate cases: one where the $\Omega_{\rm fit} = 0$ and the other where $\Omega_{\rm fit} \ne 0.$ In the first case it is possible to use the interpretation $h\approx f^{-1}\partial f/\partial t$ giving a physical interpretation to the perturbations. In the non-zero cases, the interpretation is not valid, and the non-equilibrium part is only used to sieve out acceleration effects from all of the effects. We provide the resulting non-equilibrium function in Fig.~\ref{fig:obs_h}. In the top panel, we show the $\Omega_{\rm fit}=0$ cases. One can see that the levels of non-equilibrium needed to make the data in the region self-consistent can be quite large compared to the one in the simulated galaxy (see Fig.~\ref{fig:h_amplitudes_mock}). It is also seen that the level of the detected non-equilibrium is dependent on the sample brightness limit -- the brightness limit determines the sample size and the extent of stellar orbits (see Fig.~\ref{fig:illustration_brightend} for clarification). If we do not use a bright-end cut at all, the the level of perturbations increases even more. Although both selectional effects and real non-equilibrium can produce the same outcome, we favour the real non-equilibrium since the selection effects induced by magnitude limits produce spherically symmetric over- or underestimation of densities. In contrast, a non-equilibrium population along the orbit would produce linear ones along the rotation direction. Therefore these effects are independent and distinguishable when the observer is in central parts of the region.  Overall we consider our main results based on the $\Omega_{\rm fit}=0$ modelling with less conservative selection. We consider these not being biased by brightness and having clear physical interpretation.

The derived $h$-values in cases of real MW data and of mock simulation data are different in their magnitude. For observations, the levels of non-equilibrium are larger than simulations. We used the simulation of \citet{Garbari_2011} that contained dark matter and stellar particles and no mergers nor gas particles. The lack of these two contributions can explain the excess of $h$ in observations. We can see from the Fig.~\ref{fig:obs_h} that the maximum of $h$ is around $L_z\approx-2100$, which translates to rotation speeds of the non-equilibrium orbits $\approx 255\,{\rm km\,s^{-1}}$, which is faster than the circular speed equivalence of this region ($\approx 230\,{\rm km\,s^{-1}}$). It indicates that these stars that have a non-equilibrium population along the orbits originate further away on average, and in the studied region, they are closer to their pericentre than apocentre.  An alternative interpretation would be that non-equilibrium is induced by the gas disc that rotates faster than average stellar disc (due to asymmetric drift). The disc gas co-moves with faster moving stars, and the non-equilibrium could be induced by gas distribution inhomogenities.

\section{Discussion}\label{sec:discussion}

In this paper, we demonstrate the capabilities and perspectives of inclusion of non-stationarities in modelling of the MW galaxy. First, based on our tests on simulation data, we conclude that the inclusion of non-stationarity to the modelling does not produce any additional inaccuracies to the model and helps to derive the acceleration field in perturbed states of a galaxy. 
Applying the developed modelling method to \textit{Gaia} data in SN demonstrates the ability of the method to detect perturbations, calculate their properties and possibly pinpoint their possible origin. 

The applications in the current paper studied regions of size $0.5\dots1.0\,{\rm kpc}$. We selected the small regions in order to have a minimally simple form to the acceleration. Would the regions be larger, the acceleration field should be more complex. In the mock application, the functional form is adequate to cover the gravitational potential changes (see Fig.~\ref{fig:funform_adequacy}), only in the case of radial acceleration, there is hint to a systematic trend in order to have additional acceleration term. We did not include it in present case since the systematic effects describe deviations in the order of $\pm20\potunit$, while the overall potential changes are $\sim 10\,000\potunit$. Would a dense gas disc be present then the acceleration would have more abrupt changes in the vertical direction. This would favour an additional term to a simple linear approximation.

Although selecting larger regions poses some calculation difficulties due to more complicated acceleration field form, it provides also advantages. Selecting a larger region allows orbital arcs to be longer when passing through the region, and inference allows larger number density of orbits. Since the density of orbits is larger, the causal connection between different parts of the regions is tighter, and inference of non-equilibrium ($\partial f/\partial t$) is more robust. Larger regions also pose tighter constraints on the stability of the gravitational potential, which we will dissect in the next paragraphs. 

\begin{figure}
    \centering
    \includegraphics{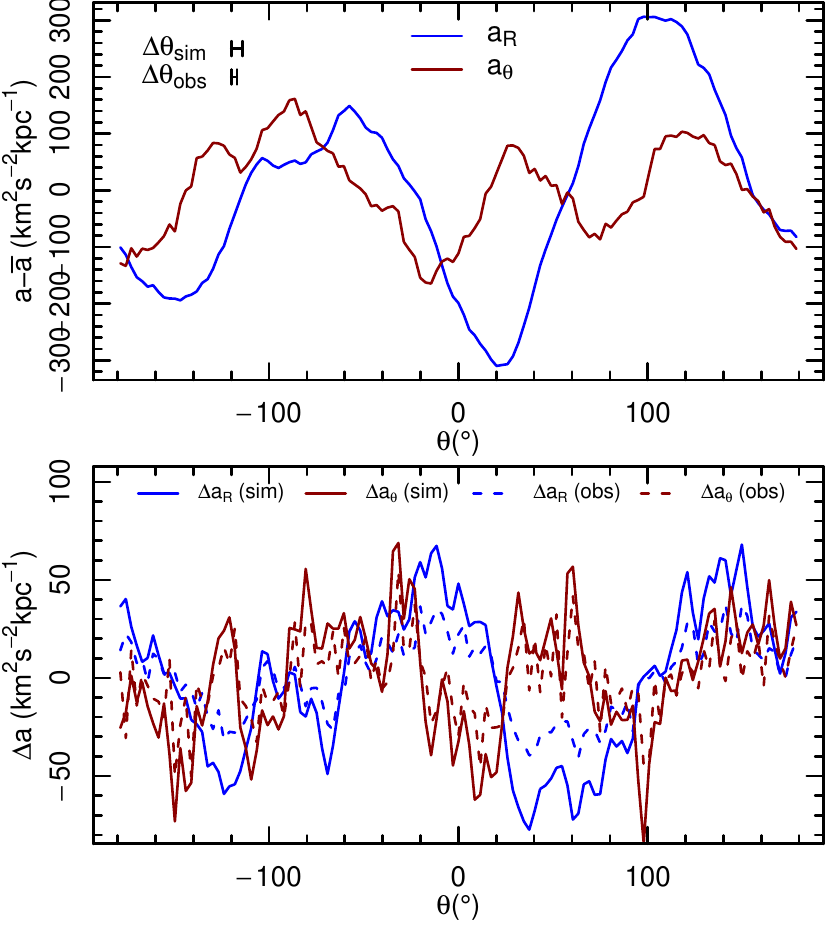}
    \caption{The position dependence of acceleration components over the simulated galaxy. The top panel shows the mean-subtracted acceleration as a function of angle measured at the centre. We can see that the acceleration components can deviate from axisymmetry up to levels $300\,{\rm km^2s^{-2}kpc^{-1}}$. The change of studied region position with respect to these asymmetries causes the potential to be time-dependent. The top left part shows the expected change of position ($\Delta \theta$) when a star passes through the studied region. The bottom panel shows the change of acceleration with the time a star passes through the studied region. It is in the order $\Delta a \lesssim 50\,{\rm km^2s^{-2}kpc^{-1}}$.}
    \label{fig:sim_pot_timedep}
\end{figure}

In principle, in a non-stationary system gravitational potential is a function of time. However, in present paper we assume that density perturbations are small and do not contain significant amount of mass. Thus, we assume that potential and acceleration fields formulae do not contain time component (see Sect.~\ref{sec:meth_general_impr}). We are modelling only non-stationarity of the density. Thus we assess now how accurate this assumption is using the simulated data. 
When calculating stellar orbits in a selected region potential changes in time because the position of the region changes with respect to the overall asymmetries of the galaxy. Figure \ref{fig:sim_pot_timedep} shows the acceleration components as a function of the position in the galaxy, more specifically, as a function of the position angle measured at $R = 8.3\,{\rm kpc}.$ We see that accelerations change by up to $\sim 300\,{\rm km^2s^{-2}kpc^{-1}}$ as a star circles the entire galaxy. As the region limits the extent of stars position range,  corresponding accelerations vary less. Let us estimate  expected acceleration changes due to change of the potential. 

The radius of the studied region is about $1\,{\rm kpc}$ and semi-height  $0.35\,{\rm kpc}$. The average rotational speed in the SRs is $\approx 210\,{\rm km\,s^{-1}}$, indicating that it takes $\Delta t\approx 3\,{\rm Myr}$ for a random star in the studied region to reach the boundary of the region. This estimate includes reducing sizes due to the flattened geometry of the region but ignores the dispersion component of the velocities.

Let us denote $\Omega_{\rm p}$ as the angular speed of the asymmetries that cause changes of acceleration. For bar, it is $\Omega_{\rm p}\approx37\,{\rm km\,s^{-1}kpc^{-1}}$ (see Sect. \ref{sec:mock_data}). The angle (measured from the centre of the galaxy) from the bar tip changes by the time an average star passes the studied region $\Delta \theta = \Omega_{\rm p} \Delta t \approx 6^\circ$. The corresponding value for the SN region of the MW is $\Delta \theta_{\rm obs} = 3^\circ$. It is smaller due to smaller extent of the SN region. The top left of the top panel of the Figure \ref{fig:sim_pot_timedep} shows these values as the lengths of horizontal bars. The bottom panel of the figure shows how much accelerations change with this rotation angle. There is a significant dependence where the studied region is located in the mock galaxy. Although there is a dependence, the order of magnitude is low compared to the mean radial acceleration. The absolute mean of the components describes how much change for the acceleration ($\Delta a$) we expect when the star passes the studied region. For simulations $\Delta a_R \approx 33$ and $\Delta a_\theta \approx 26 \,{\rm km^2s^{-2}kpc^{-1}}$, and when assuming similar acceleration distribution for MW than the simulation has, then the corresponding values are $\Delta a_R \approx 17$ and $\Delta a_\theta \approx 15 \,{\rm km^2s^{-2}kpc^{-1}}$. The reduced size of the studied region is the cause of the less sensitivity to the time dependence of the potential. These values show that the time-dependence can cause acceleration changes in the order of $33/6000\approx 0.5\%$, which we consider insignificant for this type of modelling and conclude that the orbital arc method is insensitive to the gravitational potential time variations in case of small regions of interest.

The case of consistent stationary systems density and velocity changes are inter-related and must be consistent. If the density changes cannot be modelled consistently with corresponding velocity changes, they are presumably caused by non-equilibrium. Another cause for density changes without corresponding velocity changes is due to observational selection function. Our preliminary testings show that confusing these two reasons in modelling is possible and can cause biased results.  

%---- future plans ----
\section{Conclusions}\label{sec:summary}

An application to the Solar Neighbourhood (SN) we came to the following conclusions.
\begin{itemize}
    \item The developed method gives consistent results for the radial acceleration ($-6360\accunit$, corresponding to circular velocity $228\,{\rm km\,s^{-1}}$; see Table~\ref{tab:obsresults}) compared to previous circular velocity estimates $238\,{
    \rm km\,s^{-1}}$ \citep{Schonrich:2012}. While other studies have used data spread over large regions, we estimated the same quantity based on the data within $0.5\,{\rm kpc}$, assuming Solar velocity from \citet{Malhan:2020}.
    \item When looking at the angular momentum distribution of the non-stationarity function $h(L_z)$, we see that all perturbed stars are populating faster (presumably thin disc) orbits. 
    \item When studying the angular momentum $L_z$ of the  perturbed stars     we found them to be near $L_z\approx 2100\,{\rm km\,s^{-1}\,kpc}$. As this is larger than the angular momentum corresponding to the circular speed of SN ($\approx1900\,{\rm km\,s^{-1}\,kpc}$), hence they must be nearer to their pericentre than compared to their circular speed. We propose that their origin is from larger radii. An alternative is that the stars that rotate faster are sensitive to co-moving perturbations by a component that rotates faster than the stellar disc. This source of acceleration is presumably inhomogenities of gas disc.
    \item Compared to \citet{Kipper:2020} we updated the local tangential acceleration estimate to include non-equilibrium. The tangential acceleration estimate is $-245\accunit$ (see Table~\ref{tab:obsresults}), which would indicate the updated bar mass estimate is $1.3\times10^{10}\,{\rm M_\odot}$.
\end{itemize}

In the case of precision modelling of the disc of MW with future \textit{Gaia} DR3+ data, we can describe the major source of non-stationarity nearby. In the case of modifying Eq.~\eqref{eq:Lz_pert_form} to a sum of perturbations, we can cover most sources of non-stationarity. 

As for future applications, the orbital arc method provides an invaluable addition to contemporary halo studies, where interactions are searched in action space. The action space has the sensitivity to mergers smoothed out long ago, while detection of non-stationarities provides information on freshly happening events. By the freshly happening influences, we mean the non-stationarities produced by substructure: perhaps there is a dark matter sub-halo, which produces dynamical friction and the wake alongside it. The wake by itself is a non-stationarity (not bound system) hence detectable at shorter timescales. Applying this methodology in the MW halo opens new and exciting opportunities to search for. There are also already more known areas that localised acceleration field could shed light on, such as the mass of Magellanic clouds could be determined by precision acceleration field, or exact shape of MW dark matter halo in case the observations with an excellent data quality are be made of outer halo. 

\section*{Acknowledgements}
We thank the referee for his/her thorough work on the paper and comments. 
This work was supported by institutional research funding  \mbox{PUTJD907}, and \mbox{PRG1006} of the Estonian Ministry of Education and Research. RK thanks FINCA and Turku University for hosting me for the postdoc studies. We acknowledge the support by the Centre of Excellence "The Dark Side of the Universe" (TK133) financed by the European Union through the European Regional Development Fund. 
This work has made use of data from the European Space Agency (ESA) mission Gaia (https://www.cosmos.esa.int/gaia), processed by the Gaia Data Processing and Analysis Consortium (DPAC, https://www.cosmos.esa.int/web/gaia/dpac/consortium). Funding for the DPAC has been provided by national institutions, in particular the institutions participating in the Gaia Multilateral Agreement.
\bibliographystyle{mnras}

\section*{Data availability statement}
The data is freely acquired from Gaia DR2 web page, and StarHorse project page \citep{starhorse}. The code used for the analysis and posterior distribution of the modelling can be acquired by contacting the corresponding author of the paper. 

\bsp	% typesetting comment
\label{lastpage}
\end{document}